\documentclass[preprint,amsmath,amssymb,superscriptaddress,notitlepage]{revtex4-1}
\usepackage{graphicx}
\usepackage[utf8]{inputenc}
\usepackage{amssymb}
\usepackage{color}
\usepackage{units}
\usepackage{float}
\usepackage{amsmath}
\usepackage{url}
\usepackage{hyperref}
\usepackage{epsfig}
\usepackage{bm}
\hypersetup{
  colorlinks  = true, 
  urlcolor   = blue,
  linkcolor  = red, 
  citecolor  = blue 
}

\begin{document}
\title{Large Nernst effect in Te-based van der Waals materials}
\author{M. Behnami}
\affiliation{IFW Dresden, P.O. Box 270116, 01171 Dresden, Germany}
\affiliation{Institut f{\"u}r Festk{\"o}rper- und Materialphysik, Technische Universit{\"a}t Dresden, 01062 Dresden, Germany}
\affiliation{Department of Physics, University of Genoa, 16146 Genova, Italy}

\author{M. Gillig}
\affiliation{IFW Dresden, P.O. Box 270116, 01171 Dresden, Germany}

\author{A. G. Moghaddam}
\affiliation{Department of Physics, Institute for Advanced Studies in Basic Sciences (IASBS), Zanjan 45137-66731, Iran}
\affiliation{Computational Physics Laboratory, Physics Unit, Faculty of Engineering and Natural Sciences, Tampere University, FI-33014 Tampere, Finland}

\author{D. V. Efremov}
\affiliation{IFW Dresden, P.O. Box 270116, 01171 Dresden, Germany}

\author{G. Shipunov}
\affiliation{IFW Dresden, P.O. Box 270116, 01171 Dresden, Germany}

\author{B. R. Piening}
\affiliation{IFW Dresden, P.O. Box 270116, 01171 Dresden, Germany}

\author{I. V. Morozov}
\affiliation{IFW Dresden, P.O. Box 270116, 01171 Dresden, Germany}

\author{S. Aswartham}
\affiliation{IFW Dresden, P.O. Box 270116, 01171 Dresden, Germany}

\author{J. Dufouleur}
\affiliation{IFW Dresden, P.O. Box 270116, 01171 Dresden, Germany}

\author{K. Ochkan}
\affiliation{IFW Dresden, P.O. Box 270116, 01171 Dresden, Germany}

\author{J. Zemen}
\affiliation{Institute of Physics ASCR, v.v.i., Cukrovarnick\'a 10, 162 53, Praha 6, Czech Republic}

\author{V. Kocsis}
\affiliation{IFW Dresden, P.O. Box 270116, 01171 Dresden, Germany}

\author{C. Hess}
\affiliation{IFW Dresden, P.O. Box 270116, 01171 Dresden, Germany}
\affiliation{Fakultät für Mathematik und Naturwissenschaften, Bergische Universität Wuppertal, 42097 Wuppertal, Germany}

\author{M. Putti}
\affiliation{Department of Physics, University of Genoa, 16146 Genova, Italy}
\affiliation{CNR-SPIN, 16152 Genova, Italy}

\author{B. Büchner}
\affiliation{IFW Dresden, P.O. Box 270116, 01171 Dresden, Germany}
\affiliation{Institut f{\"u}r Festk{\"o}rper- und Materialphysik, Technische Universit{\"a}t Dresden, 01062 Dresden, Germany}

\author{F. Caglieris}
\affiliation{CNR-SPIN, 16152 Genova, Italy}

\author{H. Reichlova}
\affiliation{IFW Dresden, P.O. Box 270116, 01171 Dresden, Germany}
\affiliation{Institut f{\"u}r Festk{\"o}rper- und Materialphysik, Technische Universit{\"a}t Dresden, 01062 Dresden, Germany}
\affiliation{Institute of Physics ASCR, v.v.i., Cukrovarnick\'a 10, 162 53, Praha 6, Czech Republic}

\date{\today}

\begin{abstract}

Layered van der Waals tellurides reveal topologically non-trivial properties that give rise to unconventional magneto-transport phenomena. Additionally, their semimetallic character with high mobility makes them promising candidates for large magneto-thermoelectric effects. Remarkable studies on the very large and unconventional Nernst effect in WTe$_2$ have been reported, raising questions about whether this property is shared across the entire family of van der Waals tellurides.

In this study, systematic measurements of the Nernst effect in telluride van der Waals Weyl semimetals are presented. Large linear Nernst coefficients in WTe$_2$ and MoTe$_2$ are identified, and moderate Nernst coefficients with non-linear behavior in magnetic fields are observed in W$_{0.65}$Mo$_{0.35}$Te$_2$, TaIrTe$_4$, and TaRhTe$_4$. Within this sample set, a correlation between the dominant linear-in-magnetic-field component of the Nernst coefficient and mobility is established, aligning with the established Nernst scaling framework, though with a different scaling factor compared to existing literature. This enhancement might be caused by the shared favorable electronic band structure of this family of materials.
Conversely, the non-linear component of the Nernst effect in a magnetic field could not be correlated with mobility. This non-linear term is almost absent in the binary compounds, suggesting a multiband origin and strong compensation between electron-like and hole-like carriers. This comprehensive study highlights the potential of van der Waals tellurides for thermoelectric conversion.

\bigskip

\end{abstract}

\maketitle
\section{Introduction}
Transition metal ditellurides WTe$_2$ and MoTe$_2$ exhibit a plethora of fascinating physical properties. They are layered van der Waals (vdW) materials, as illustrated in Figure 1a, formed by 2D-planes weakly connected by vdW forces and characterized by minimal element intermixing \cite{ajayan2016van, novoselov20162d, geim2013van, dorrell2020van}.

Such a peculiar structure allows for efficient exfoliation into thin flakes with atomically smooth interfaces, enabling effective modification of their electronic properties by tuning their dimensionality down to the 2D limit. This makes them highly attractive in the fields of nanoelectronics and quantum devices. Among their most appealing characteristics, WTe$_2$ hosts quantum spin Hall states both in bulk and monolayer forms and exhibits giant non-saturating magnetoresistance \cite{tang2017quantum, song2018observation, ali2014large}. Additionally, unconventional superconducting quantum criticality has been reported in monolayers of WTe$_2$ \cite{song2024unconventional}. Furthermore, the orthorhombic ($T_d$) phases of both WTe$_2$ and MoTe$_2$ have been identified, both theoretically and experimentally, as inversion-symmetry-breaking type-II Weyl semimetals  \cite{soluyanov2015type, li2017evidence, bruno2016observation, wu2016observation, sun2015prediction, deng2016experimental, jiang2017signature}. Weyl semimetals host relativistic Weyl fermions at topologically protected band-touching points characterized by linear dispersion relations \cite{yan2017topological, jia2016weyl, armitage2018weyl, sakhya2023observation}. The topological protection makes Weyl nodes robust against small perturbations, and due to their defined chirality, they act as sources and sinks of Berry curvature \cite{wuttke2019berry}. As a result, unconventional transport properties have been reported in MoTe$_2$ and WTe$_2$ \cite{li2017evidence, kononov2020one, kang2019nonlinear, wu2018observation, liang2019origin, ma2022growth, chen2016extremely}.


 The rich phenomenology observed in $T_d$-MoTe$_2$ and WTe$_2$ combined with their 2D layered nature inspired an intense search for similar structures and led to the synthesis of the relative tellurides TaIrTe$_4$ and TaRhTe$_4$ \cite{koepernik2016tairte,haubold2017experimental, shipunov2021layered}. In these compounds, W is replaced by Ta and Ir or Rh, which form a zigzag chain along the $a$-direction \cite{shipunov2021layered}, as shown in Figure 1a. In analogy to WTe$_2$ also these compounds crystallize in orthorhombic structure, belonging to the non-centrosymmetric space group (No: Pmn2$_1$). Thanks to this structure, a minimal admitted number of four Weyl points has been predicted and consequently confirmed by angle-resolved photoemission spectroscopy (ARPES) \cite{koepernik2016tairte, haubold2017experimental}. The characteristic features of their band structure \cite{shipunov2021layered} shared with the parental WTe$_2$ \cite{soluyanov2015type} include the proximity of tilted Weyl cones to the Fermi level and almost compensated electron-like and hole-like bands as schematically illustrated in Figure 1b. Remarkably, the unique non-trivial electronic band structure of TaIrTe$_4$ lead to observation of nonlinear Hall effect, quantum spin Hall effect, dual quantum spin Hall insulator or spin-orbit torque \cite{kumar2021room, cheng2024giant, guo2020quantum, tang2024dual, bainsla2024large}, and TaRhTe$_4$ is equally promising \cite{zhang2024layer}.\\
 

Similar to other semimetals with high mobilities \cite{wularge, yamada2024nernst}, vdW tellurides are promising candidates for efficient thermoelectric transport. The Nernst effect, which generates a voltage perpendicular to both the temperature gradient and an external magnetic field (as illustrated in Figure 1c), is particularly interesting because it enables thermoelectric devices to be composed of a single material, unlike Seebeck modules that require two different materials \cite{feng2024transverse}. The Nernst effect is odd in a magnetic field, meaning the direction of the generated electric field can be controlled by the direction of the applied magnetic field. Furthermore, hybrid structures and multilayers are being explored to enhance and control the Nernst effect and its anomalous counterpart \cite{uchida2024hybrid, zhou2021seebeck, boona2016observation}. VdW materials are exceptionally suitable for this research direction due to their ability to exfoliate and stack with other materials \cite{pasquale2024electrically} and to be twisted \cite{li2024dynamical} to engineer the Nernst response. A very high Nernst coefficient at low temperatures has been reported in WTe$_2$ \cite{pan2022ultrahigh, zhu2015quantum}. It has been attributed to a combination of high mobility, low effective masses of the Fermi pockets, and almost perfect compensation of electron-like and hole-like charge carriers avoiding Sondheimer cancellation \cite{sondheimer1948theory, bel2003ambipolar, sharma2017nernst}, as schematically shown in Fig.1b. Besides the exceptional magnitude, the most surprising feature of the Nernst effect in WTe$_2$ is its linearity as a function of the applied magnetic field up to 9 T. In fact, considering reported mobilities of the order of $\mu \approx$ 10$^5$ cm$^2$/Vs, the low field limit ($\mu$B$<$1), in which a linear Nernst effect is expected according to the semiclassical theory, should be valid approximately up to B $\approx$ 0.1 T. In exfoliated flakes of WTe$_2$, on the other hand, a non-linear Nernst is observed and discussed in the context of Weyl physics \cite{rana2018thermopower}. These pioneering reports in WTe$_2$ motivate a systematic study of the Nernst effect across the family of vdW tellurides.


\begin{figure}[H]
	\centering
	\includegraphics[width=\textwidth]{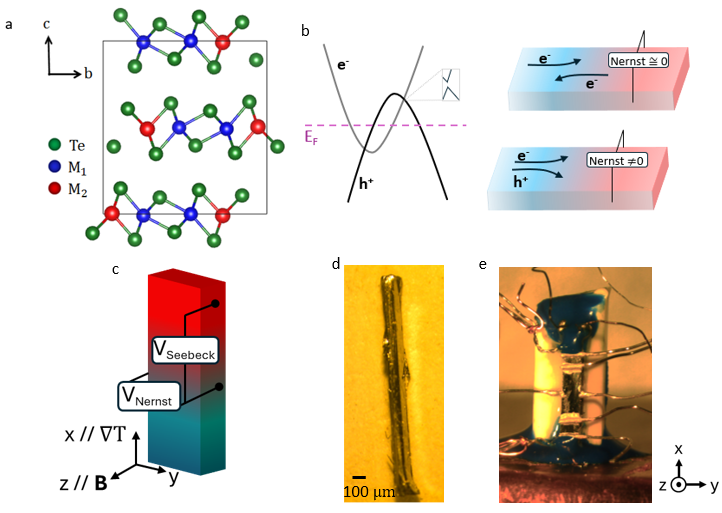}
	\caption{(a) Shared structure of the studied member of the Te-based family. The parental compound WTe$_2$ is an atomically layered transition metal dichalcogenide representing the minimal time-reversal symmetry-breaking model. The zig-zag character shown in plane c can be maintained by replacing the W atoms with other elements and combining two elements leading to the sample series presented here. (b) Illustration of characteristic features: tilted Weyl cones with favorable linear dispersion and electron-like and hole-like band compensation. Compared to a simple single-band picture where the Nernst effect is absent, the multi-band scenario enhances the Nernst response because the hole and electrons will move in opposite directions. (c) Schematic illustration of Nernst effect measurement geometry. (d) A microscope photo of our single crystal of TaRhTe$_4$. (e) Photo of a typical sample mounting to measure the magneto-thermal transport in our samples.\\
\label{f1-1}}
\end{figure}



In this work, we investigate the Nernst effect in single crystals of WTe$_2$, MoTe$_2$, W$_{0.65}$Mo$_{0.35}$Te$_2$, TaIrTe$_4$, and TaRhTe$_4$. We chose these materials because they share the same crystal structure. The band structure calculation of the WTe$_2$, MoTe$_2$, TaIrTe$_4$, and TaRhTe$_4$ can be found elsewhere \cite{di2017three, guguchia2017signatures, shipunov2021layered}. Our systematic approach allows us to identify common properties across this family of tellurides. We observe an unexpected scaling of the Nernst coefficient with the ratio of mobility to Fermi energy ($\mu / E_F$), which significantly deviates from the established Nernst scaling picture. Additionally, deviations from the linear field dependence of the Nernst effect are observed in W$_{0.65}$Mo$_{0.35}$Te$_2$, TaIrTe$_4$, and TaRhTe$_4$, but are absent in the binary compounds WTe$_2$ and MoTe$_2$. Finally, we also verify that the large Nernst response in one compound - TaIrTe$_4$ - also persists in a device fabricated from an exfoliated flake. Our study suggests that the characteristic features of van der Waals telluride compounds—specifically, the compensation of electrons and holes and the proximity of Weyl cones to the Fermi level—are the origins of the enhanced linear Nernst effect.



\section{Experimental Results}

The single crystals are grown using the self-flux method, as detailed in \cite{naidyuk2019yanson,sykora2020disorder,pawlik2018thickness,shipunov2021layered}. One example of a studied crystal of TaRhTe$_4$ crystal is shown in Figure 1d. Subsequently, surfaces contaminated with a small amount of flux are mechanically cleaved to prepare the crystals for further studies.

The samples are prepared for DC electric transport measurements by attaching silver wires using conductive silver paint. For the thermoelectric measurements, the samples are mounted in a vertical orientation on a thermally insulating plastic block as shown in Figure 1e. A resistive heater is mounted to one side of the samples using thermally conductive glue, while the opposite side is linked to the thermal mass of our probe. Temperature differences are monitored using a calibrated Chromel-Au-Chromel thermocouple, and thermoelectric signals are collected via silver wires serving as electrodes, detailed further in the supplementary information \cite{SIMahdi}.

All measurements are conducted using custom-built probes, which are inserted into Oxford cryostats equipped with 15 T magnets. Given that the Nernst (N) effect is antisymmetric with respect to the magnetic field, 
$N$ is derived by antisymmetrizing the measured signals at both positive and negative magnetic fields to eliminate potential spurious contributions \cite{SIMahdi}. 

\begin{figure}[H]
	\centering
	\includegraphics[width=\textwidth]{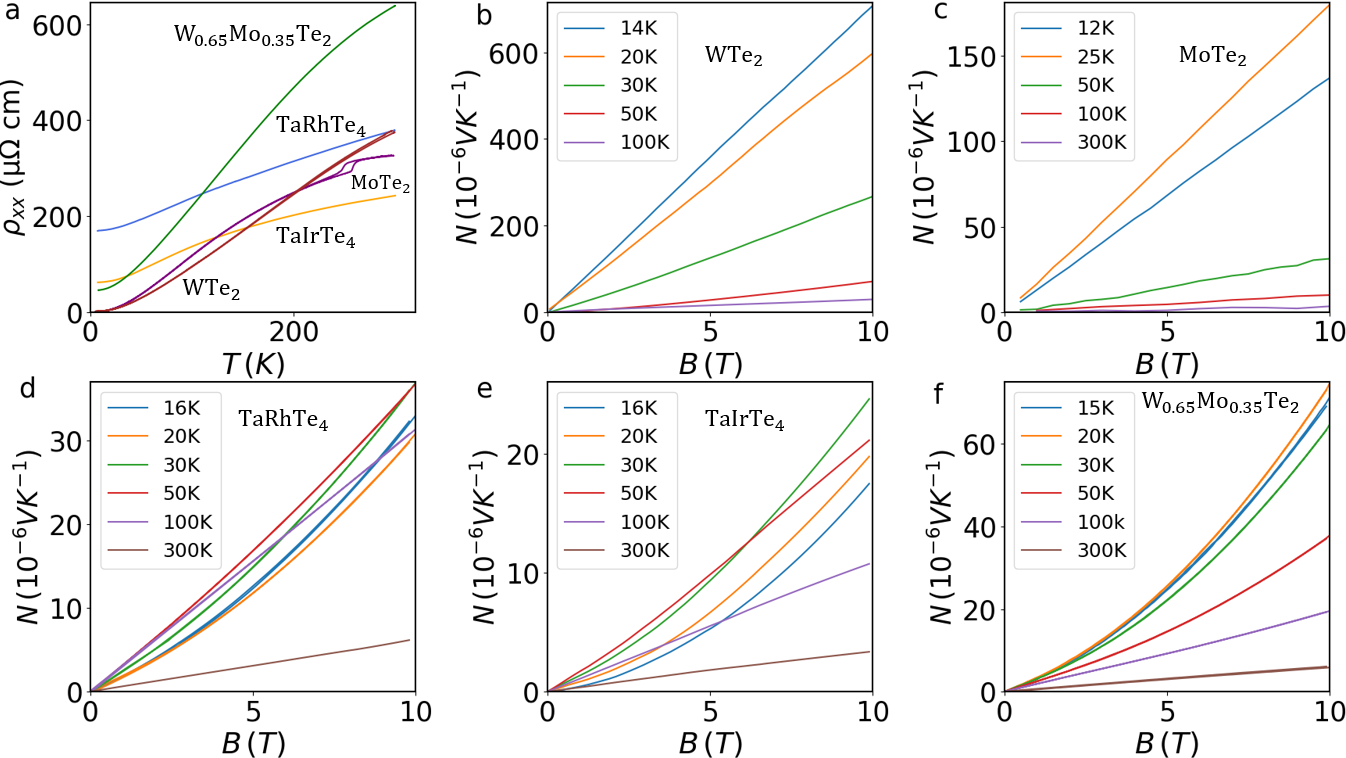}
	\caption{(a) Temperature-dependent resistivity for all studied samples. The MoTe$_2$ exhibits structural transitions around 250 K \cite{paultopological}. (b) Nernst coefficients as a function of magnetic field for WTe$_2$ which exhibit a linear behavior for all temperatures and the maximum value for the Nernst coefficient is 700 $\mu V/K$ at T=14 K and B=10 T. (c) Nernst coefficients as a function of magnetic field for MoTe$_2$ which exhibit a linear behavior (similar to WTe$_2$) and the maximum value for the Nernst coefficient is 180 $\mu V/K$ at T=25 K and B=10 T. (d) Nernst coefficients as a function of magnetic field for TaRhTe$_4$ which exhibit a non-linearity in low temperatures and linearity in high temperatures and the maximum value for the Nernst coefficient is 37 $\mu V/K$ at T=30 K and B=10 T. (e) Nernst coefficients as a function of magnetic field for TaIrTe$_4$ which exhibit a non-linearity in low temperatures and linearity in high temperatures (similar to TaRhTe$_4$) and the maximum value for Nernst coefficient is 25 $\mu V/K$ at T=30 K and B=10 T. (f) Nernst coefficients as a function of magnetic field for W$_{0.65}$Mo$_{0.35}$Te$_2$ which exhibit a very clear non-linearity in low temperatures and linearity in high temperatures and the maximum value for Nernst coefficient is 75 $\mu V/K$ at T=20 K and B=10 T.\\
\label{f2-1}}
\end{figure}

Figure 2a shows the temperature dependence of the electrical resistivity $\rho_{xx}$ for all the compounds. All the samples exhibit a metallic behavior, with $\rho_{xx}$ decreasing with the temperature and residual resistivity ratios (RRR = $\rho_{xx}$(300~K)/$\rho_{xx}$(8~K)) are  RRR= 189, 183, 14, 3.9, 2.2 for MoTe$_2$, WTe$_2$, W$_{0.65}$Mo$_{0.35}$Te$_2$, TaIrTe$_4$, and TaRhTe$_4$, respectively. This suggests an overall increase of the disorder in the ternary compounds. This is also in agreement with the presence of Shubnikov-de Haas oscillations only for the binary compounds (shown in supplementary information Fig. S3 and S4 \cite{SIMahdi}). Figure (\ref{f2-1}) b-f shows a magnetic field dependence of the Nernst coefficient at various temperatures for all the studied compounds. In WTe$_2$ (Figure \ref{f2-1}b) and MoTe$_2$ (Figure \ref{f2-1}c) $N$ is linear in $B$ across the whole temperature range. In contrast, the sample characterized by a partial intermixing of W and Mo, W$_{0.65}$Mo$_{0.35}$Te$_2$, shows a clear deviation from the linear trend when the temperature is reduced below 100 K. A similar trend is observed in  TaRhTe$_4$ and TaIrTe$_4$ compounds, where a linear Nernst coefficient persists at 300 K and 100 K, but a non-linear contribution emerges at lower temperatures. Concerning the absolute value, $N$ it is very large in WTe$_2$, reaching about 600 $\mu$V/K at $T= 20$ K and $B= 10$ T. Interestingly, $N$ progressively decreases in the other compounds, achieving 180 $\mu$V/K, 75 $\mu$V/K, 37 $\mu$V/K and 25 $\mu$V/K in MoTe$_2$, W$_{0.65}$Mo$_{0.35}$Te$_2$, TaRhTe$_4$ and TaIrTe$_4$, respectively.

\section{Discussion}
A peculiar characteristic of some of our materials is the appearance of a non-linear Nernst effect as a function of the magnetic field. Typically, a deviation from the $B$-linear standard Nernst coefficient is interpreted in the framework of the Anomalous Nernst Effect (ANE), which could have an intrinsic origin, driven by a non-vanishing Berry curvature \cite{cheng2024tunable, ceccardi2023anomalous, ding2019intrinsic, zeugner2023weyl, yang2020giant, roychowdhury2023anomalous}, or an extrinsic one, depending on the skew scattering or the side-jump scattering mechanisms by magnetic impurities \cite{wularge, caglieris2018anomalous, liang2017anomalous}. However, the characteristic phenomenology of the ANE is a tendency to saturation of $N$ to a constant value with increasing the magnetic field, while in W$_{0.65}$Mo$_{0.35}$Te$_2$, TaRhTe$_4$ and TaIrTe$_4$, in the low-temperature region below 100 K, the $N$ vs $B$ curves assume an unusual super linear trend.
In order to analyze our data we opted for a phenomenological model, which, by symmetry considerations (the Nernst coefficient is odd in $B$), includes an additional cubic term summed to the standard linear one, namely $N=$ a$B$+c$B^3$. The fitting results are generally satisfactory in the whole temperature range for all the investigated materials (shown in supplementary information Fig. S5 \cite{SIMahdi}). 

$N$ can be decomposed into the difference between the tangent of the thermoelectric angle $\frac{\alpha_{xy}}{\alpha_{xx}}$ and the tangent of the Hall angle $\frac{\sigma_{xy}}{\sigma_{xx}}$, multiplied by the Seebeck coefficient $S$:
\begin{align}
 N=S\left(\frac{\alpha_{xy}}{\alpha_{xx}}-\frac{\sigma_{xy}}{\sigma_{xx}}\right),
\label{equ3}
\end{align}
where $\sigma_{xy}=\frac{\rho_{xy}}{\rho_{xx}^2+\rho_{xy}^2}$ and $\sigma_{xx}=\frac{\rho_{xx}}{\rho_{xx}^2+\rho_{xy}^2}$ \cite{yang2023anomalous,pu2008mott}. We evaluated $S\frac{\sigma_{xy}}{\sigma_{xx}}$, observing that it negligibly contributes to $N$ for all of our compounds (shown in supplementary information Fig. S6 \cite{SIMahdi}). Hence, in our case, the observed $B^3$ term must be determined by the thermoelectric angle and in particular by the $B$-antisymmetric term $\alpha_{xy}$ \cite{yang2023anomalous,pu2008mott}.
Remarkably, equation \eqref{equ3} still holds in multiband materials.

In materials with a strong compensation between electron-like and hole-like carriers, it is immediate to demonstrate that a $B^3$ contribution can emerge in $N$. By assuming a two-carrier-type model in which, the mobility $\mu$ is the same for holes and electrons, the most simplified solution for the Boltzmann equation, leads to the following expression up to the third order (further details in the supplementary information  \cite{SIMahdi}):

\begin{align}
N
 &\approx
  \frac{\pi^2}{6} \: \frac{k_B}{e} \: k_BT 
   \left\{
2\mu B\frac{\partial_\varepsilon(\sigma^{0}_e\sigma^{0}_h)}{(\sigma^{0})^2}\: 
\left[1-(x_{\sigma}\mu B)^2\right]
+(\partial_\varepsilon \mu)B \left[1-(\mu B)^2(2-x_\sigma^2)\right]
 \right\}
 + {\cal O} (B^5),
 \end{align}

 where $\sigma_e$ and $\sigma_h$ are the electrical conductivity of the electron- and hole-like band, respectively, $\sigma^0=\sigma_e+\sigma_h$, $x_\sigma=(\sigma_e-\sigma_h)/(\sigma_e+\sigma_h)$ and the energy derivative $\partial_\epsilon$ is evaluated at the Fermi energy $E_F$. In our materials, it turns out that the linear and the cubic terms of $N$ have the same positive sign. According to Eq. 2, this condition is verified if, for instance, $\partial_\epsilon(\sigma_e\sigma_h)$ and $\partial_\epsilon\mu$ have different signs. This happens if the mobility decreases with $E_F$ (i.e. due to an increment of the scattering), while the overall conductivity increases due to an enhancement of the carrier density, which overcomes the mobility attenuation. 

The electron-hole compensation therefore makes the cubic term more pronounced. However, its amplitude also depends on the mobility and its energy dependency, which are influenced by the material's band structure as well as extrinsic parameters or temperature. This could explain why we do not observe non-linearity in the Nernst measurements in WTe$_2$.

Turning to the linear contribution, the conventional Nernst effect is described by the following formula in the picture of an electron fluid $N/B = \frac{\pi^2}{3}\frac{k_B}{e}\frac{k_B T}{E_F}\mu$. Here, $k_B$ is the Boltzmann constant, $e$ is the elementary charge of an electron, $E_F$ is the Fermi energy, and $\mu$ is the electron mobility. According to this formula, the Nernst effect is directly proportional to the electron mobility scaled by the Fermi energy \cite{behnia2009nernst, behnia2016nernst}.

In previous reports \cite{behnia2016nernst,behnia2009nernst,yang2023anomalous}, it has been shown that this scaling law is robust for an extended variety of materials, ranging from cuprate superconductors to elemental bismuth, as shown in Figure 3b, which presents a plot of the Nernst coefficient divided by temperature and magnetic field as a function of the mobility scaled to the Fermi energy. The red line, which interpolates the majority of the materials, represents the theoretical value predicted by the above equation and the scaling factor is determined by $L\frac{e}{k_B}$, where $L$ is the Lorenz number.\\

In order to verify such scaling for our compounds, we estimated the mobility of our materials by combining the longitudinal resistivity and the Hall resistivity (shown in supplementary information Fig. S7 \cite{SIMahdi}). We have used a two-carrier model for the materials with sufficiently large magnetoresistance and, within the error bar, the estimated mobility was in agreement with the value obtained by assuming the single band model which we finally used to estimate the mobility of all our compounds. Consistent with previous reports in WTe$_2$ \cite{ali2014large, jiang2015signature}, we also estimate similar hole and electron concentration in the studied low-temperature regime (shown in supplementary information Fig. S8 \cite{SIMahdi}).

We have chosen the same experimental conditions for all samples - temperature of $T$=20 K and magnetic field at $B$=10 T. The comparison between the linear contribution of $N$ and $\mu$ shown in Figure 3a, confirms an evident correlation between these two quantities (T=30 K and B=10 T is shown in supplementary information Fig. S9 \cite{SIMahdi}). This is in contrast with the non-linear component of the Nernst signal, which cannot be correlated with mobility, as shown in supplementary information Figure S10 \cite{SIMahdi}. The reduced mobility in our WTe$_2$ samples compared to \cite{pan2022ultrahigh} is probably also the reason for the smaller Nernst signal observed here.  

The results of our evaluation are displayed in Figure 3b. We present the measured values for all our compounds as a function of mobility and Fermi energy, revealing a departure from the Fermi liquid (FL) prediction, represented by the red curve \cite{behnia2016nernst,behnia2009nernst,yang2023anomalous}. A large error bar, indicated by the size of the markers, reflects the uncertainty associated with the estimation of mobility and Fermi energy. We have used the value of the Fermi Energy ($E_F$) of most of our samples based on the Density Functional Theory reported in the literature \cite{di2017three, guguchia2017signatures, shipunov2021layered}.  Due to difficulties to calculate the $E_F$ of the off-stochiometric W$_{0.65}$Mo$_{0.35}$Te$_2$ compound we estimated the $E_F$ which would be required to match the trend displayed in Fig.3b. The estimated $E_F \sim 110\, meV$ is realistic and it is between the values in the parental compounds WTe$_2$ and MoTe$_2$ \cite{SIMahdi}. 

This systematic enhancement of the linear component of the Nernst effect in this family of materials necessitates the search for a common feature among all the studied compounds. The most plausible explanation is the favorable combination of semimetallic character with Weyl cones near the Fermi energy, a characteristic shared across all these materials. The linear dispersion of the Weyl cones contributes to high mobility, and the second key factor leading to the enhanced Nernst effect is the multiband character with significant electron-like and hole-like band compensation. As illustrated in the schematic in Fig. 1b, unlike transverse electrical transport, the transverse thermoelectric response is not expected in a single-band model due to the cancellation between the Hall and Peltier angles, which have the same sign \cite{sharma2017nernst}. However, it is enhanced when both electron-like and hole-like contributions are present, a phenomenon sometimes referred to as the ambipolar Nernst effect \cite{bel2003ambipolar}. We speculate that, in our family of materials, the assumptions underlying the derivation of the previously reported scaling law \cite{behnia2016nernst} are significantly violated. Specifically, the simplification to a single-band model and the assumption of temperatures much lower than the Fermi temperature does not hold. In addition, the scaling factor reported in \cite{behnia2016nernst} is obtained within the assumption that the dominant scattering mechanism is the elastic one. Although this approximation might be valid in our case, it is worth noting that in the presence of non-negligible electron-electron or electron-phonon scattering the scaling factor can also sensibly change. Hence, the combination of such effects could reasonably result in a systematic deviation from the previously predicted scaling trend for our materials. We note that also the high mobility WTe$_2$ studied in \cite{pan2022giant} (green marker in Fig.3b) exhibits departure from the scaling law, consistent with our observation. 

\begin{figure}[H]
	\centering
	\includegraphics[width=\textwidth]{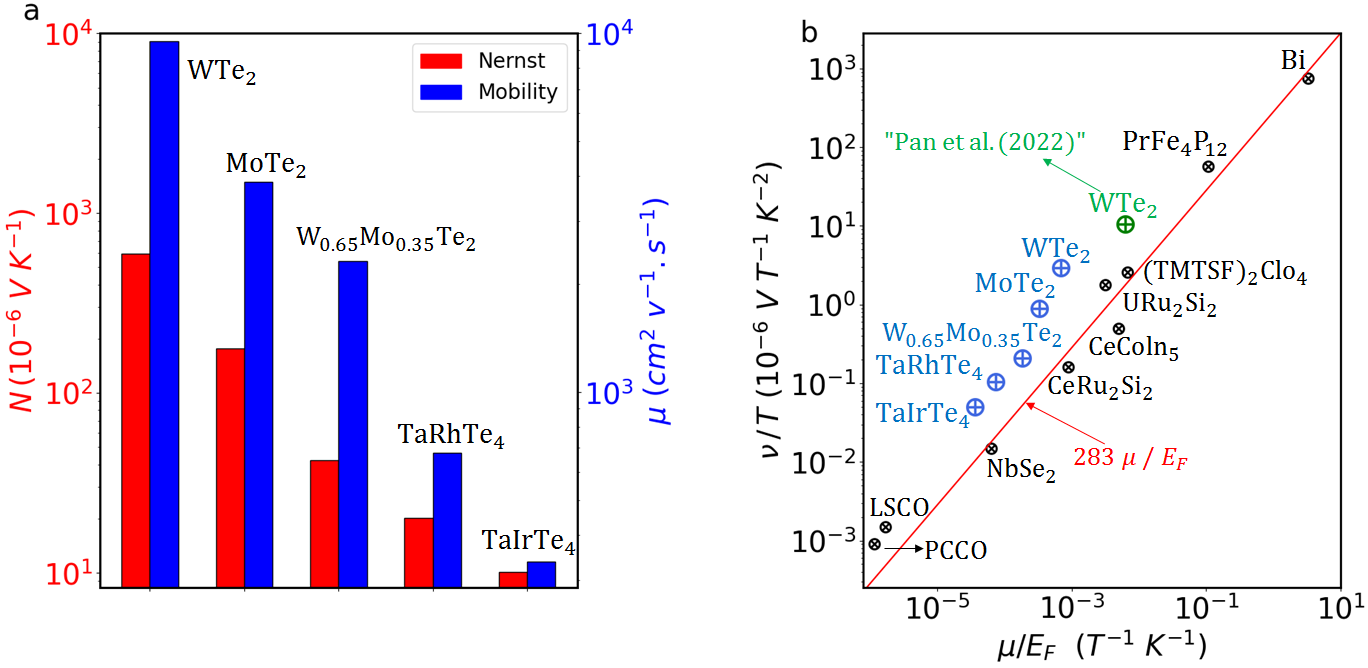}
	\caption{Comparing the linear in B contribution of the Nernst effect versus mobility at T=20 K and B=10 T for studied samples. (a) WTe$_2$ is an example of the binary compound, it has large mobility and a large linear contribution of the Nernst coefficient, and TaIrTe$_4$ is an example of the ternary compound, it has smaller mobility and a smaller linear contribution of the Nernst coefficient. (b) Overview of all measured Nernst coefficients compared to existing literature. On the y-axis is the low-temperature slope of the Nernst coefficient divided by temperature and magnetic field. On the x-axis, we plot the ratio of electron mobility to the Fermi energy. The red line corresponds to the previously reported \cite{behnia2009nernst} scaling law for the Nernst coefficient. The green marker is the value for high mobility WTe$_2$ reported in \cite{pan2022giant}. \\
\label{f3-1}}
\end{figure}

Finally, to demonstrate the robustness of the measured Nernst response, we tested its presence on an exfoliated flake of one of the studied family members, TaIrTe$_4$. In this device, we mechanically exfoliated a 20 nm thick flake onto a substrate. Electrical contacts were defined using e-beam lithography, as illustrated in Figure 4a. We used current-induced Joule heating as a source of the thermal gradient. The second harmonic voltage was employed as a sensitive measure of the thermoelectric response \cite{reichlova2015current}. The magnitude of the thermal gradient was modeled using COMSOL, as illustrated in Figures 4b and 4c using the  parameters detailed in the supplementary information \cite{SIMahdi}. The resulting thermal gradient develops mostly along the $x$-direction and it reaches 9.4~K/mm at the position of voltage detection (15~$\mu$m from contacts 1 and 2). The voltage measured at the contacts 3 and 4 is shown in Figure 4d. The evaluated Nernst signal measured at 4~K is approximately 6$\mu V/K$  at 2~T, which is consistent with the TaIrTe$_4$ bulk crystal measurement at 16~K, although the direct comparison is not possible due to slightly different temperatures on the sample in different setups. 

\begin{figure}[H]
	\centering
	\includegraphics[width=\textwidth]{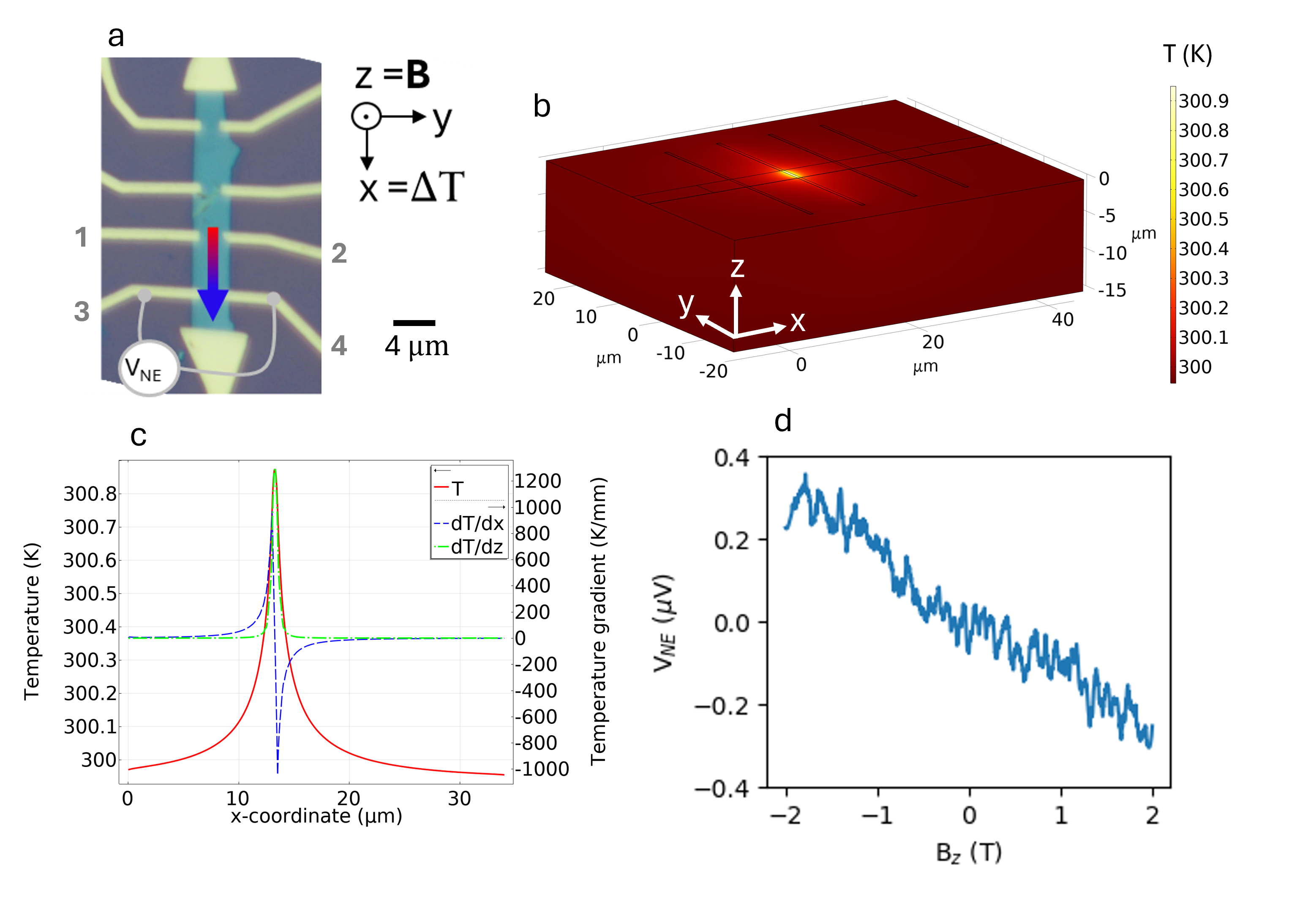}
	\caption{{ Nernst effect in exfoliated flake of TaIrTe$_4$. (a) Microscope image of an exfoliated flake with indicated experimental geometry.  AC 0.2 mA current is applied between contacts 1 and 2 and it generates a thermal gradient via Joule heating. The thermal gradient propagates along the flake in the $x$-direction. The Nernst voltage is detected on contacts 3 and 4 as the second harmonic voltage signal. (b) COMSOL simulations of the temperature distribution in the exfoliated flake. (c) Simulated thermal gradient showing value dT/dx of 9.4 K/mm at the position of contacts 3 and 4 plotted along a cut line through the center of the flake. (d) Measured Nernst voltage by the second harmonic technique}.\\
\label{f4-1}}
\end{figure}
\section{Summary and Conclusions}
In summary, we have systematically measured the Nernst effect in single crystals of tellurium-based van der Waals materials, specifically WTe$_2$, MoTe$_2$, W$_{0.65}$Mo$_{0.35}$Te$_2$, TaIrTe$_4$ and TaRhTe$_4$. We report a very large Nernst effect in the binary compounds and moderately large effects in the ternary compounds. We correlate the results with the materials' electronic mobility and observe a new scaling factor of the Nernst coefficient with $\mu / E_F$, which significantly differs from the Nernst scaling prediction
\cite{behnia2016nernst,behnia2009nernst,yang2023anomalous}. 

Furthermore, we show that the Nernst effect is also present in devices made from exfoliated flakes which highlights the robustness of the effect and the potential of this family of materials for spin-caloritronic devices. 

Additionally, we observed a non-linear contribution to the Nernst effect in the ternary compounds at low temperatures. This phenomenon could have multiple origins, including strong compensation between electron-like and hole-like carriers, topological Weyl crossings near the Fermi energy, or impurity band effects leading to multiband effects. These findings warrant further investigation. In this direction, it would be very interesting to understand the eventual role of the topological non-trivial bands by tuning the system dimensionality by applying uniaxial strain, gate voltage, or systematic doping series in order to modify the Fermi level position.

\medskip
\textbf{Supporting Information} \par 
Supporting Information is available from the Online Library or from the author\cite{SIMahdi}.

\medskip
\textbf{Acknowledgements} \par 
We would like to thank Tino Schreiner and Danny Baumann for their technical support. MB acknowledges the Graduate Academy of TU Dresden (PSP-Element: F-010000-702-3B1-2330000). SA acknowledges DFG via project number 523/4-1. SA, DE, and BB acknowledge DFG via project number 405940956. HR was supported by the Grant Agency of the Czech Republic Grant No. 22-17899K, TERAFIT - CZ.02.01.01/00/22$\_$008/0004594 and the Dioscuri Program of MPI and MEYS LV23025. VK was supported by the Alexander von Humboldt Foundation.

\medskip
\textbf{Conﬂict of Interest}\par
The authors declare no conﬂict of interest.

\medskip
\textbf{Author Contributions}\par 
B.B., H.R., and M.B. proposed the study. G.Sh., I.V.M., B.R.P., and S.A. prepared and characterized the samples. M.B., H.R., M.G., V.K., C.H., M.P., and F.C. designed the measurement setup, performed the experiments, and analyzed the data. A.G.M. and D.V.E. proposed the theoretical part. K.O. and J.D. performed the exfoliated flake measurement. J.Z. performed COMSOL simulations. B.B. supervised the study. M.B., H.R., and F.C.  wrote the manuscript with input from all authors.

\medskip
\textbf{Data Availability Statement}\par
The data that support the findings of this study are available from the
corresponding authors upon request.


\medskip

\end{document}